\documentclass[a4paper]{jpconf}
\usepackage{graphicx}

\usepackage{epsfig}
\def\lsim{\raise0.3ex\hbox{$<$\kern-0.75em\raise-1.1ex\hbox{$\sim$}}}
\def\gsim{\raise0.3ex\hbox{$>$\kern-0.75em\raise-1.1ex\hbox{$\sim$}}}
\begin{document}
\title{Lattice Simulations of the Thermodynamics of Strongly Interacting
Elementary Particles and the Exploration of New Phases of
Matter in Relativistic Heavy Ion Collisions
}

\author{Frithjof Karsch}

\address{Physics Department, Brookhaven National Laboratory,
Upton, NY 11973, USA}

\ead{karsch@bnl.gov}

\begin{abstract}
At high temperatures or densities matter formed by strongly interacting 
elementary particles (hadronic matter) is expected to undergo a  
transition to a new form of matter - the quark gluon plasma - in which
elementary particles (quarks and gluons) are no longer confined inside 
hadrons but are free to propagate in a thermal medium much larger in 
extent than the typical size of a hadron.
The transition to this new form of matter as well as properties
of the plasma phase are studied in large scale numerical calculations
based on the theory of strong interactions - Quantum Chromo Dynamics (QCD).
Experimentally properties of hot and dense elementary particle matter are 
studied in relativistic heavy ion collisions such as those currently 
performed at the relativistic heavy ion collider (RHIC) at BNL.

\noindent
We review here recent results from studies of thermodynamic properties of 
strongly interacting elementary particle matter performed on 
Teraflops-Computer. We present results on the QCD equation of state and 
discuss the 
status of studies of the phase diagram at non-vanishing baryon number density.
\end{abstract}


\vspace*{-15.0cm}
\hfill BNL-NT-06/26 

\hfill August 2006
\vspace*{14.2cm}

\section{Introduction}

During recent years our faith in numerical calculations of properties
of strongly interacting matter at high temperature and non-vanishing
baryon number density greatly increased. The steady improvement of discretization 
schemes for the fermion sector of QCD and the development of new simulation
algorithms \cite{rhmc} now
allow to perform calculations with greatly reduced systematic 
errors. Moreover, the widespread availability of a new generation of 
Teraflops-Computer for lattice gauge theory calculations now allows to 
perform studies of thermal properties of matter formed by strongly
interacting elementary particles with an almost realistic quark
mass spectrum \cite{aoki,Bernard,schmidteos}. 

\begin{figure}
  \includegraphics[width=18pc]{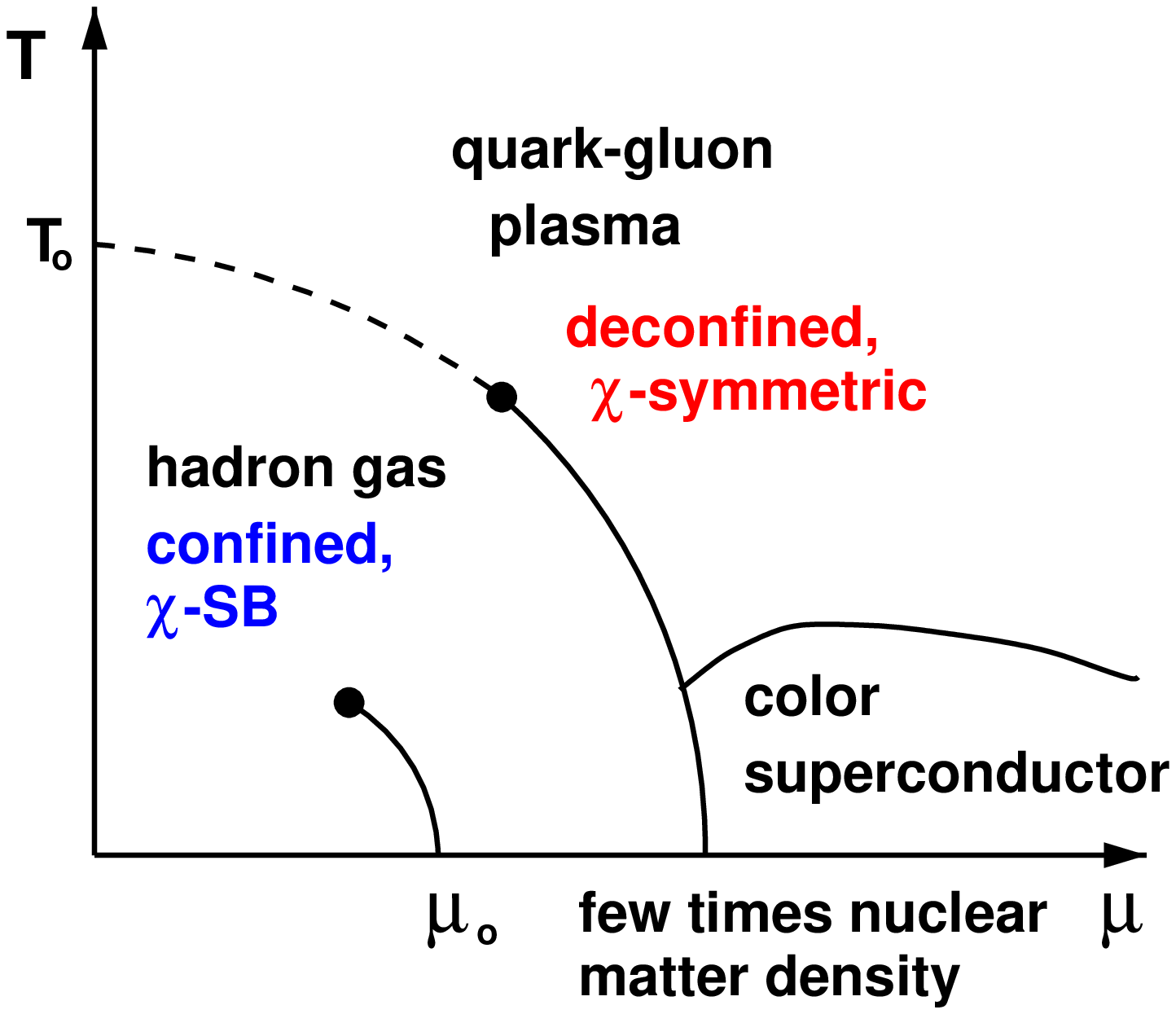}\hspace{2pc}
\begin{minipage}[b]{18pc}
  \includegraphics[width=16pc]{T_muB_gsi2.eps}
\end{minipage}
\caption{\label{fig:phasediagram}Sketch of the QCD phase diagram (left)
as well as the freeze-out curve determined   
in heavy ion collisions at various center of mass energies (right) 
\cite{cleymans}.
The freeze-out curve is parametrized by
$\displaystyle{T= 0.17 - 0.13 \mu_B^2 - 0.06 \mu_B^4}$.
}
\end{figure}

Numerical calculations with small systematic errors also form the basis for 
further quantitative studies of the QCD phase diagram at non-zero quark 
chemical potential ($\mu_q$) \cite{Fodor1,us1,Gavai1,Lombardo1,Philipsen1}
or, equivalently, non-zero baryon number \cite{redlich,deForcrand}. 
The different approaches 
developed for this purpose are still limited to the regime of  
temperatures close to and above the transition temperature, $T_0$, as well as
small values of the chemical potential, 
$T\; \gsim\; 0.9T_0,~\mu_q/T\; \lsim\; 1$. As such they do not yet allow to 
explore the interesting low temperature and high density part of the 
QCD phase diagram, where various color superconducting phases are 
expected to show up \cite{rajagopal,shovkovy}. They, however, allow  
to study the phase diagram and thermodynamic properties of matter 
in a regime accessible to heavy ion experiments and cover a regime
that will be studied in future experiments planned at RHIC and the GSI
in Germany to explore matter at high baryon number density. 
The generic form of the QCD phase diagram is shown in 
Fig.~\ref{fig:phasediagram}(left). 
Various model calculations suggest that at low temperature 
the low and high density regions in the QCD phase diagram
are separated by a line of first order phase transitions.
On the other hand, lattice calculations suggest that at small values 
of $\mu_q/T$ the 
transition from low to high temperature is not a phase transition; 
thermodynamic quantities like the energy density or the chiral condensate 
change smoothly, although quite rapidly, in a narrow temperature interval.
It thus has been speculated \cite{Stephanov} that a $2^{nd}$ order phase 
transition point exists somewhere in the interior of the QCD phase diagram.
Lattice calculations have provided first indications for the existence
of such a critical point \cite{Fodor1,Fodor2,gavai}.
Its exploration and the detailed quantitative analysis of transition
parameters characterizing the separation line between the low and high
density regime are currently being performed in large scale numerical 
simulation of QCD.   

Experimentally properties of hot and dense matter are studied 
in relativistic heavy ion collisions performed at CERN and BNL. 
In particular, recent results from RHIC experiments on on jet 
modifications and flow properties of matter created in Au-Au collisions
suggest that in the collision of two heavy nuclei a dense medium
is generated which equilibrates quickly after the collision at temperatures
well above the transition temperature estimated in lattice calculations. 
A hot and dense system thus seems to be generated in the plasma phase 
of QCD; once equilibrated it expands isentropically and cools
down again. Its constituents, quarks and gluons, then 'freeze-out',
{\it i.e.} recombine to ordinary hadrons that become experimentally 
detectable. The relative abundances of various hadron species at 
freeze-out seem to be well described by a hadron resonance gas which
allows to  relate the observed particle yields to temperature and
baryon chemical potential ($\mu_B=3\mu_q$) at freeze-out. The freeze-out parameters 
extracted from heavy ion experiments performed with different collision energies
\cite{cleymans} are shown in Fig.~\ref{fig:phasediagram}(right). 
At least for small values of $\mu_q$ the freeze-out temperature seems to 
be close to the transition temperature determined in lattice 
calculations \cite{PHENIX,STAR}. To quantify this agreement and the 
relation between freeze-out conditions and phase transitions also at 
larger values of $\mu_q$ is one of the challenges
for experimental and theoretical studies of QCD thermodynamics. 
 
We will start our survey of lattice calculations of QCD thermodynamics
in the next Section by discussing recent studies of the QCD 
equation of state  at vanishing chemical potential. 
In Section 3 we discuss the extension of these calculations to non-zero
chemical potential. We conclude in Section 4.

\section{The QCD equation of state at vanishing baryon number density}

Most information on the structure of the high temperature phase of QCD and 
the nature of the transition itself has been obtained through lattice 
calculations performed in the limit of vanishing baryon number density or, 
equivalently, vanishing quark chemical potential ($\mu_q=0$). 
This limit is most relevant for
our understanding of the evolution of the early universe. It also corresponds
to the regime which currently is studied experimentally in heavy ion 
collisions at RHIC (BNL) and soon will be explored also at the LHC (CERN). 
The experimental accessibility of this regime of dense matter also asks
for a thorough quantitative study of the QCD phase transition and of
basic parameters that characterize the thermodynamics of dense 
matter at high temperature, e.g. the transition temperature $T_0$ and
the energy density $\epsilon_c$ at this temperature. 
Good quantitative control over the temperature dependence of basic quantities
that characterize bulk properties of a thermal medium also is needed to 
extract the equation of state, $p(\epsilon)$, which controls the
evolution of dense matter created in heavy ion collisions. We will in
the following present some of the recent results on the QCD transition
and the equation of state obtained in large scale numerical simulations of 
lattice regularized QCD.

\begin{figure}[t]
\begin{center}
\epsfig{file=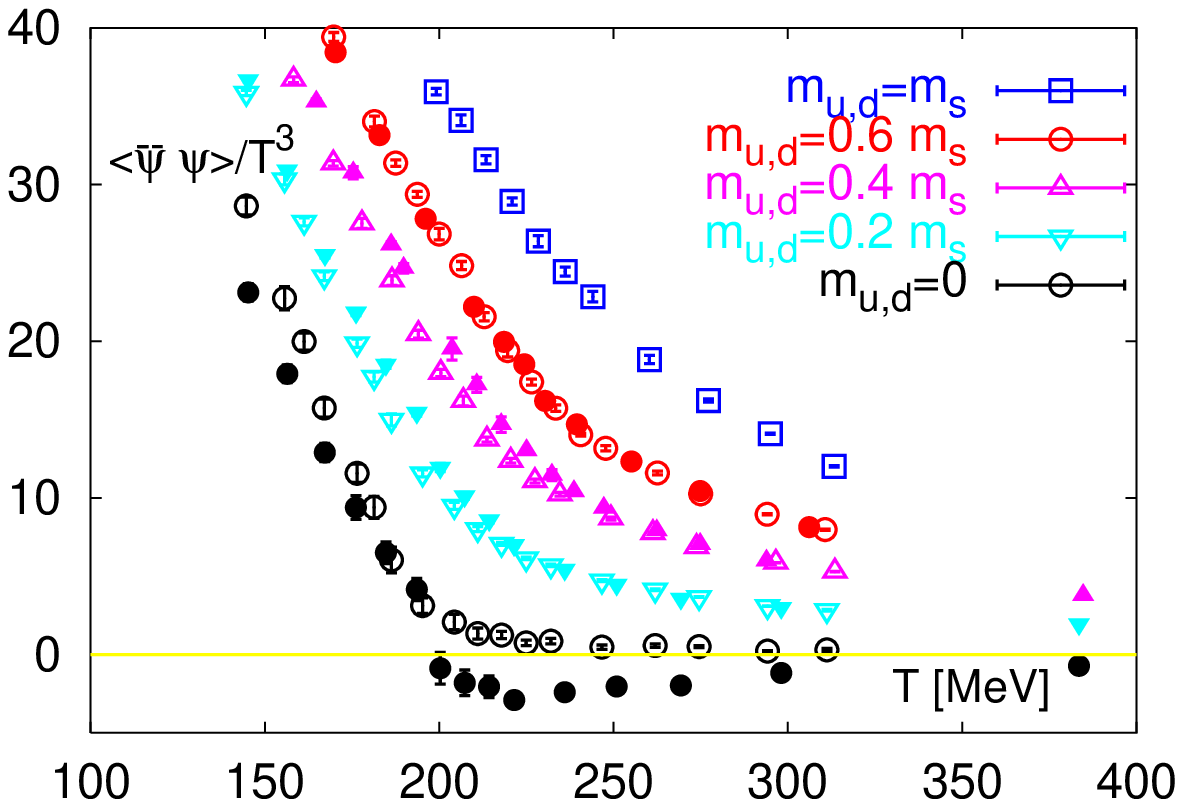,width=74mm}
\epsfig{file=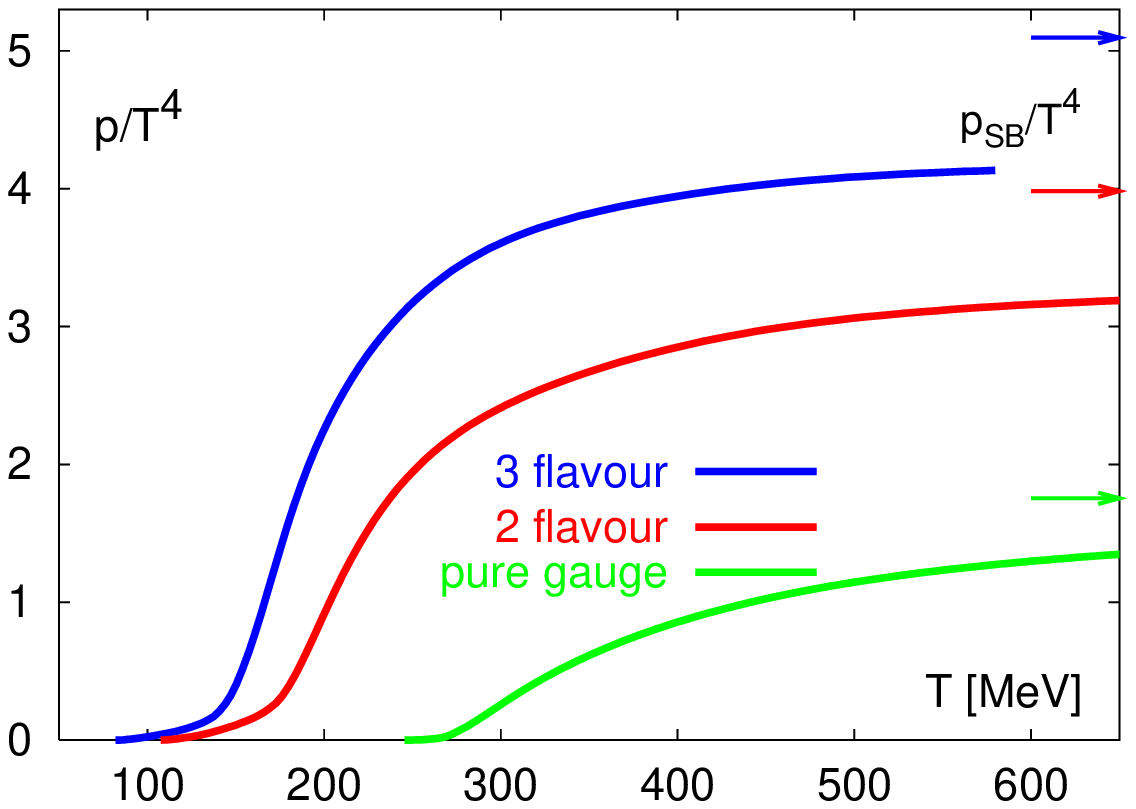,width=74mm}
\end{center}
\caption{\label{fig:chiral} The light quark chiral condensate in QCD
with 2 light up, down and a heavier strange quark mass (open symbols)
and in 3-flavor QCD with degenerate quark masses
(full symbols) \cite{Bernard04}. The right hand part of the figure shows the
pressure calculated in QCD with different number of flavors as well as
in a pure gauge theory \cite{peikert,Boyd}. 
}
\end{figure}

\subsection{Deconfinement and chiral symmetry restoration}

The transition to the high temperature phase of QCD is closely related
to two fundamental properties of QCD - chiral symmetry breaking and 
confinement. 
In QCD at low temperature chiral symmetry is spontaneously broken;
the chiral condensate, obtained as the derivative of the logarithm of
the QCD partition function, $Z(T,V,m_q)$ with respect to the quark mass,
\begin{equation}
\langle \bar\psi\psi\rangle = \frac{T}{V}\frac{\partial \ln Z}{\partial m_q}
\; ,
\end{equation}
remains non-zero even in the limit of vanishing quark mass $m_q$. In this
limit the chiral condensate is an order parameter for the phase transition 
to the high temperature phase; it is non-zero at low temperature and  
vanishes for temperatures larger
than the transition temperature. Lattice calculations give evidence for
chiral symmetry restoration as is apparent from Fig.~\ref{fig:chiral}(left) 
which shows a recent analysis of $\langle \bar\psi\psi\rangle$
performed for QCD with two light and a heavier strange quark mass as well
as for QCD with three degenerate quark masses \cite{Bernard04}. Although the 
transition is not a true phase transition but a smooth crossover, it clearly 
is signaled by a rapid change in the chiral condensate. 

The chiral
transition also is deconfining, {\it i.e.} at the transition temperature
a large number of new degrees of freedom gets liberated. These degrees
of freedom can be identified as quarks and gluons. At (very) high 
temperatures bulk thermodynamic observables resemble the behavior of an
ideal quark-gluon gas, {\it i.e} the pressure
and energy density approach the corresponding Stefan-Boltzmann values,
\begin{equation}
{p_{\rm SB} \over T^4} = {1\over 3} {\epsilon_{\rm SB} \over T^4}
={\pi^2 \over 45}\biggl( 8 +{21\over 4} n_f \biggr) \; .
\end{equation}
This becomes apparent when one compares, for instance, results for the 
temperature dependence of the pressure calculated in QCD with different 
number of quark species ($n_f$ flavors).
As can be seen in Fig.~\ref{fig:chiral}(right) the pressure rises rapidly
at the transition temperature and at higher temperatures approaches (slowly)
the Stefan-Boltzmann limit for an ideal gas with the relevant number of 
quark and gluon degrees of freedom.

\subsection{The QCD transition temperature}

\begin{figure}[t]
\begin{center}
\epsfig{file=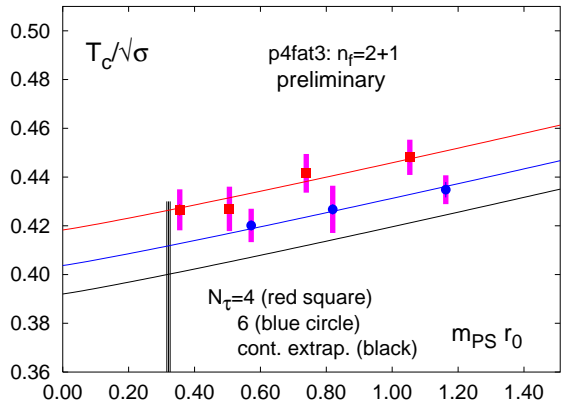,width=82mm}
\end{center}
\caption{\label{fig:Tc} The transition temperature in QCD expressed in
units of the square root of the string tension. Shown are results obtained
from a simulation of QCD with two light quark masses and a heavier strange
quark mass \cite{RBC-BI}. They are plotted versus the lightest pseudo-scalar 
meson mass expressed in terms of the commonly used scale parameter $r_0$ which
is determined from the slope of the heavy quark potential at short
distances (for more details on the scale $r_0$ see for instance 
\cite{Bernard04}).} 
\end{figure}

Determining the transition temperature accurately is one of the basic
goals of numerical studies of QCD thermodynamics. This involves several 
independent steps. 
First of all one has to determine the bare coupling, $g^2$,
at which the transition takes place, on a four dimensional lattice with
fixed extent, $N_\tau$, in the fourth direction. This determines the
transition temperature in units of the lattice spacing $T=1/N_\tau a(g^2)$.
In a second step one has to determine the lattice spacing, $a(g^2)$, that 
corresponds
to this value of the gauge coupling. This can be done by calculating 
another physical observable in lattice units that is known experimentally, 
e.g. a hadron mass, $m_Ha(g^2)$, or a phenomenologically known observable like
the heavy quark potential. Finally one has to increase the lattice 
extent $N_\tau$ and reduce the quark masses to perform a controlled
extrapolation to the continuum limit with physical values of the quark 
masses. In Fig.~\ref{fig:Tc} we show results from an ongoing analysis 
\cite{RBC-BI} that follows this program. Shown there is
the transition temperature expressed in terms of the 
square root of the string tension, $\sqrt{\sigma}$, which characterizes
the long distance behavior of the heavy quark potential, 
\begin{equation}
V_{q\bar{q}}(r) = - \frac{\alpha}{r} + \sigma r \; . 
\label{scales}
\end{equation}
Similar studies with almost physical light quark masses and a heavier
strange quark mass have been performed recently by the MILC 
collaboration \cite{Bernard04}. 
These analyses have been performed with significantly smaller quark masses
and a smaller lattice spacing than earlier studies. They indicate that 
the transition temperature expressed in units of the string tension,
$T_c/\sqrt{\sigma}\simeq 0.39-0.4$, is
in fact somewhat smaller than earlier estimates. Nonetheless,
recent studies of heavy quarkonium spectra, suggest a value for the string
tension, $\sqrt{\sigma}\simeq 460$~MeV, which is about 10\% larger than
values used previously to convert the QCD transition temperature to
physical units. This leads to somewhat larger estimates for $T_0$ than
the value $T_0\simeq 175$~MeV used in the past. 
Current estimates suggest a transition temperature
of about $190$~MeV for QCD with almost physical light quark masses and
a heavier strange quark mass. 

\subsection{The QCD equation of state at $\mu_q=0$}

As has been discussed in Section $2.1$ we learn from the
temperature dependence of the equation of state about the relevant
degrees of freedom in the high temperature phase of QCD. 
Reaching accurate quantitative results on the temperature dependence
of pressure and energy density also is needed for the comparison
of experimental results obtained in heavy ion collisions with 
theoretical calculations performed in equilibrium QCD. The dense
system created in a heavy ion collision rapidly expands and cools down
after its generation and equilibration. This expansion process may 
be described using hydrodynamic models. In these models the equation
of state is a basic input. 
  
\begin{figure}
  \includegraphics[height=.24\textheight]{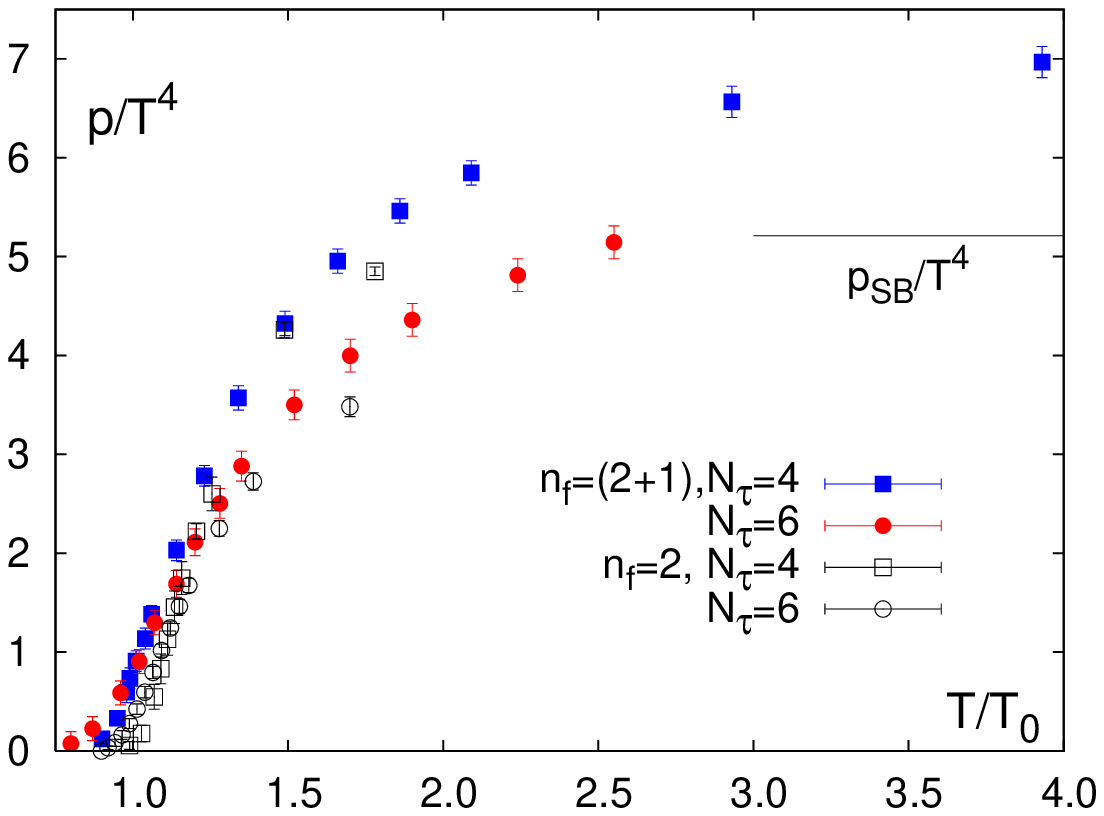}
  \includegraphics[height=.23\textheight]{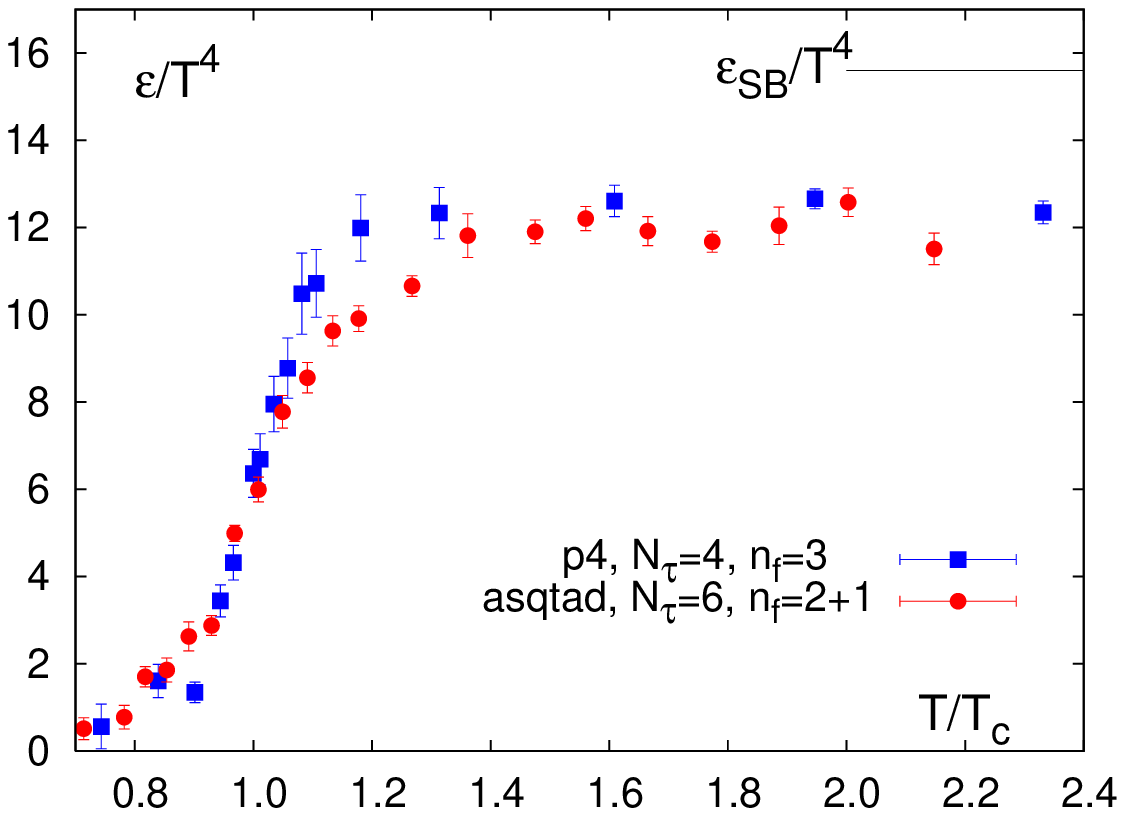}
  \caption{Cut-off dependence of the pressure calculated with the standard 
staggered fermion action on lattices with temporal extent $N_\tau =4$ and
6 in 2-flavor QCD \cite{milc_nt4,milc_nt6} and (2+1)-flavor QCD \cite{aoki} 
(left).  The right hand figure shows the energy density 
calculated with improved staggered fermion actions. Shown are
results for 3-flavor QCD obtained 
with the so-called p4-action on lattices with temporal extent
$N_\tau =4$ \cite{peikert} and for (2+1)-flavor QCD obtained
with the asqtad-action for $N_\tau =6$ \cite{Bernard}.
}
\label{fig:pressure}
\end{figure}

While earlier studies of the equation of state of QCD have been performed with
rather heavy quark masses and large lattice spacings, recent studies 
significantly improved over this situation and allow to get better control
over systematic effects arising from the use of unphysically large quark mass
values as well as from the use of too coarse lattices.
Although the calculations with light quarks have still been performed
in rather small physical volumes, $TV^{1/3}\simeq 2$, they do support earlier 
findings on the temperature dependence of the pressure and energy density
in the transition region and also confirm that 
quark mass effects are small at high temperature. 
In particular, they show that the contribution of strange quarks, which
have a mass of the order of the transition temperature, $T_0$,
has little influence on the thermodynamics
in the vicinity of $T_0$. In Fig.~\ref{fig:pressure}(left) we compare the
recent calculation of the pressure in (2+1)-flavor QCD \cite{aoki}
and earlier results for 2-flavor QCD \cite{milc_nt4,milc_nt6} which both
have been performed with unimproved staggered fermions.
The good agreement in the vicinity of $T_0$ suggests
that the strange quark contribution to the pressure is small. Only for 
$T\; \gsim \; 1.5 T_0$ differences show up; the positive strange quark 
contribution to the pressure in $(2+1)$-flavor
QCD becomes sizeable.

In Fig.~\ref{fig:pressure}(right) we compare results for the energy density
calculated in QCD with two light quark masses and a heavier strange quark
mass \cite{Bernard} with results obtained in 3-flavor QCD with a 
quark mass that rises with temperature but stays small on the
scale given by the temperature, $m/T < 1$ \cite{peikert}. 
The good agreement between these calculations is quite reassuring. 
These calculations confirm that thermodynamics in the high 
temperature phase 
is rather insensitive to changes of the quark mass;
a reduction of the light quark masses by almost an order of
magnitude does not lead to drastic changes in the energy density at the
transition point and in the high temperature phase.
These recent calculations also
show that the transition itself is not strongly influenced by
discretization errors, which in the staggered fermion formulation
show up prominently in the distortion of the light hadron spectrum; 
reducing $m_q$ and thus the masses of light hadrons as 
well as reducing flavor symmetry breaking effects drastically \cite{aoki} 
does not 
significantly change the energy density at the transition temperature.
The estimate, $\epsilon_c/T_0^4 = 6\pm 2$ \cite{peikert}, is consistent
with the recent calculations in (2+1)-flavor QCD performed with lighter
quark masses. The weak dependence of transition parameters on the quark
mass can be understood in phenomenological models of the low temperature
phase of QCD; it seems to 
reflect the importance of numerous heavy resonances that are necessary
to build up the particle and energy density needed for the 
transition to occur \cite{redlich_HG}.

Like earlier calculations with improved staggered fermions also the recent 
studies of the equation of state performed at vanishing quark chemical 
potential 
suggest that for physical values of the quark masses the transition to the
high temperature phase of QCD only is a rapid crossover rather than a phase
transition which on finite lattices would be signaled by metastabilities
and a strong volume dependence of bulk thermodynamic observables or the 
chiral condensate. 
None of the calculations performed so far for QCD with two light
quarks with or without the inclusion of a heavier strange quark gave direct 
evidence for a first order phase transition. 

\section{Thermodynamics at non-zero baryon number density}

\subsection{Isentropic equation of state and freeze-out in heavy
ion collisions}

Studies of the QCD equation of state have recently been extended to the
case of non-zero quark chemical potential ($\mu_q$). Calculations of bulk
thermodynamic quantities for $\mu_q > 0$
based on the reweighting approach \cite{Fodoreos},
using the Taylor expansion
of the partition function \cite{Allton2,Allton4,isentropic}, 
as well as analytic continuation of calculations performed with
imaginary values of the chemical potential \cite{lombardo}
show that the 
$\mu_q$-dependent contributions to energy density and pressure are
dominated by the leading order $(\mu_q/T)^2$ correction. For RHIC energies,
$\mu_q/T\simeq 0.1$, even this contribution is negligible.

\begin{figure}
\begin{center}
\epsfig{file=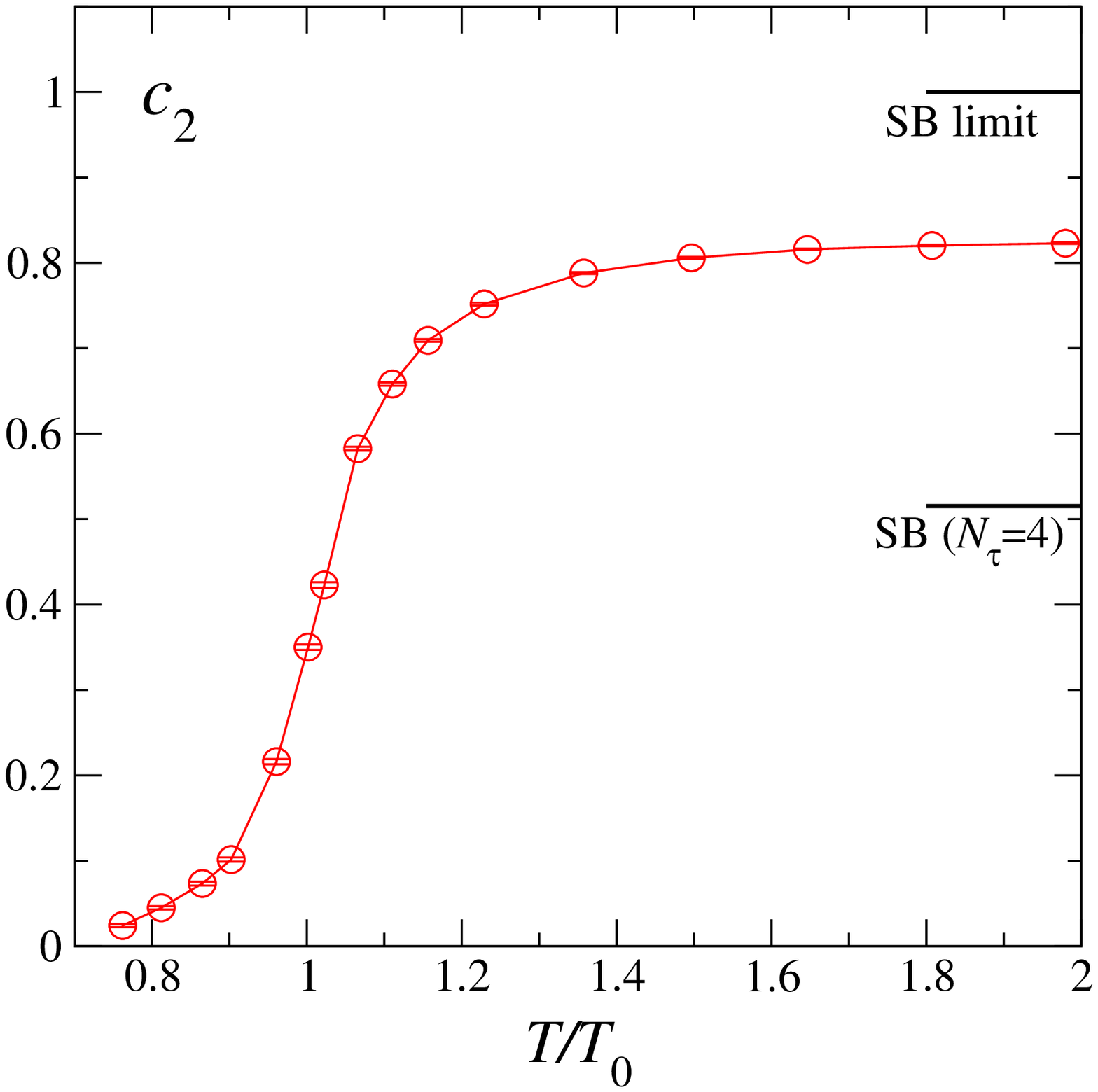,width=4.7cm}
\epsfig{file=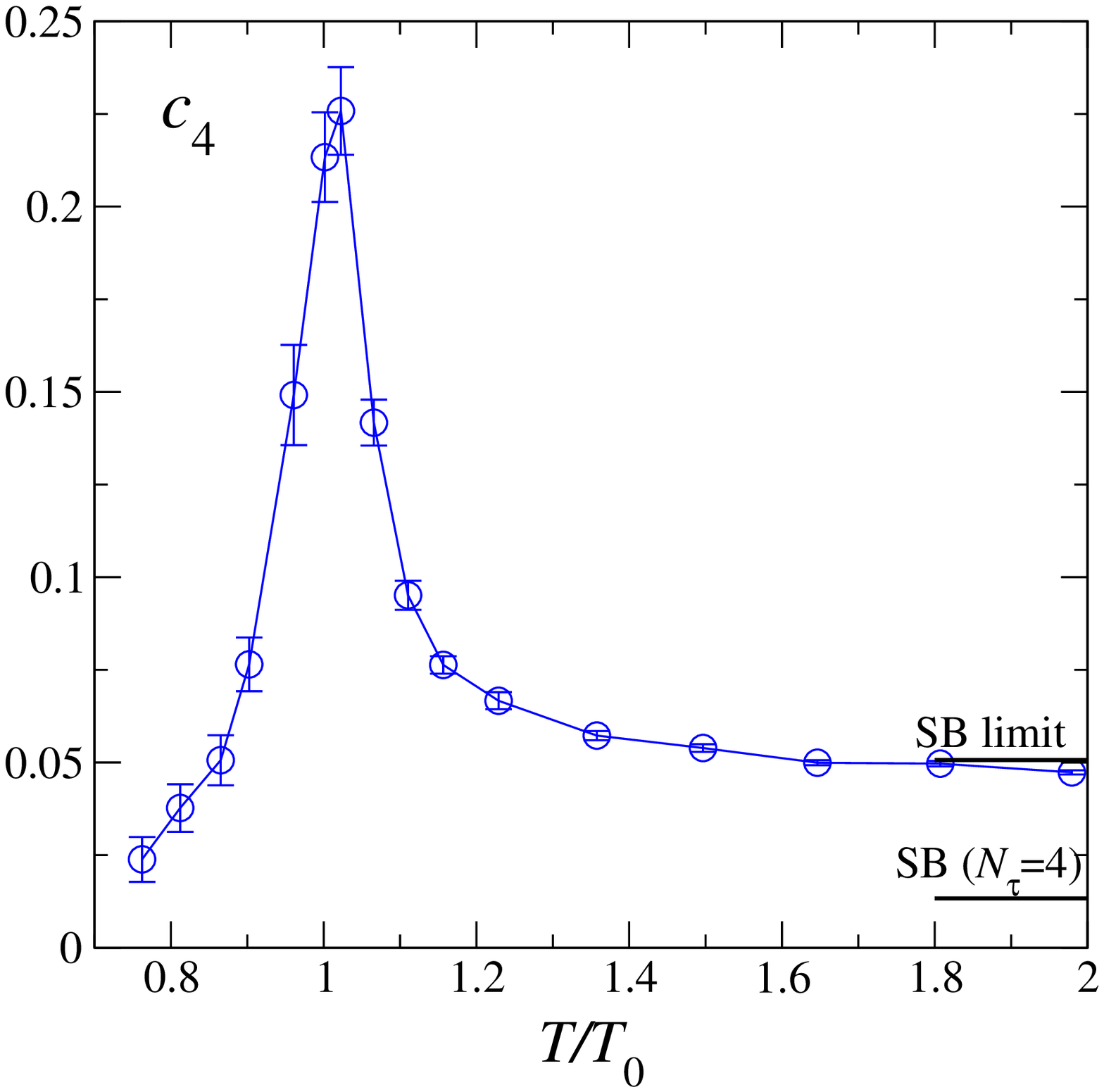,width=4.7cm}
\epsfig{file=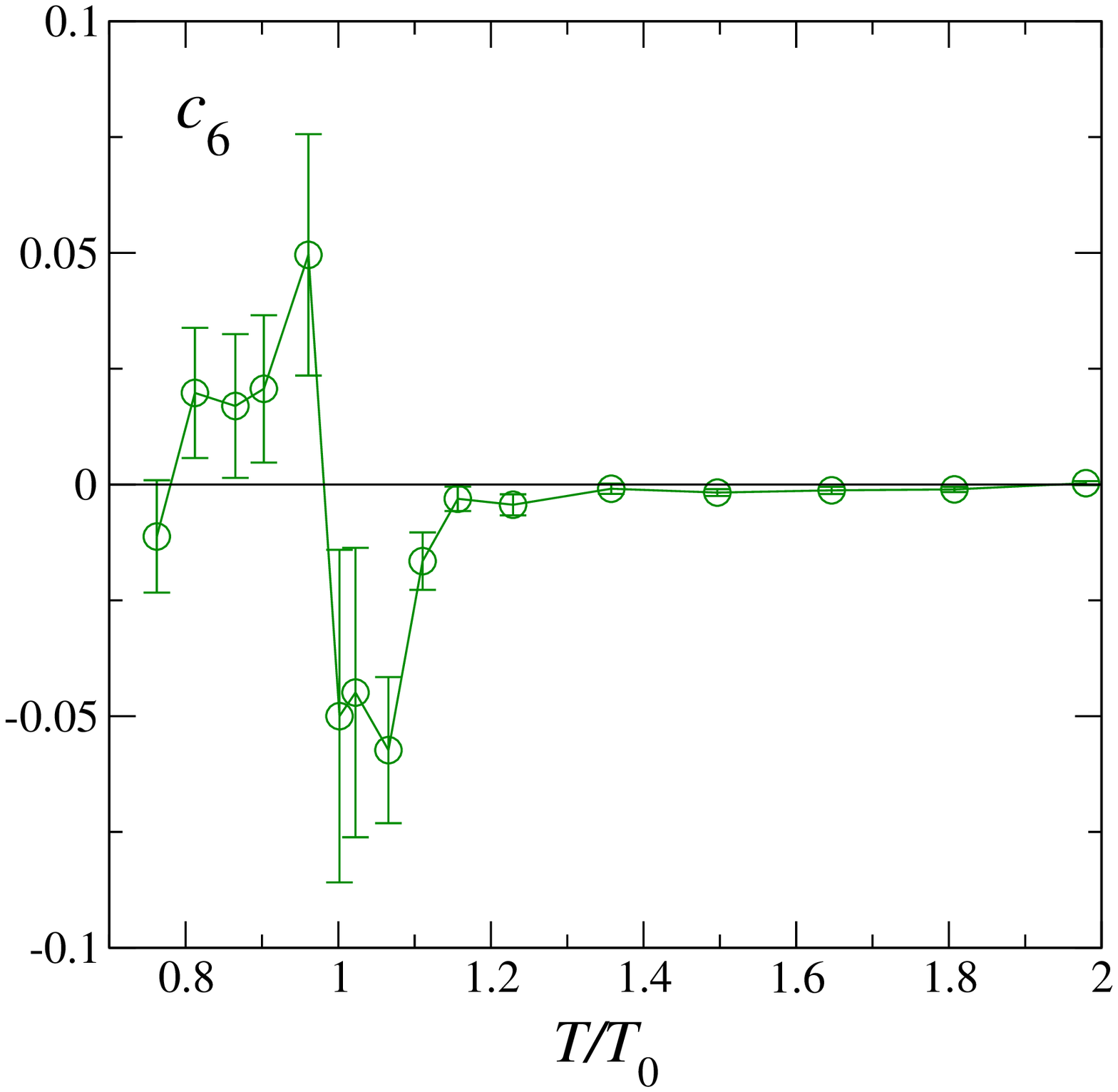,width=4.7cm}
\end{center}
\caption{\label{fig:coefficients} Temperature dependence of the Taylor
expansion coefficients for $p/T^4$
in 2-flavor QCD and for quark masses corresponding at $T_0$ to a pseudo-scalar 
(pion) mass of about $770$~MeV.
}
\end{figure}

We will focus in the following
on a discussion of Taylor expansions of the partition function of
2-flavor QCD around $\mu_q=0$.
At fixed temperature and small values of the chemical potential the pressure
may be expanded in a Taylor series around $\mu_q = 0$,
\begin{equation}
{p\over T^4}={1\over{VT^3}}\ln Z = \sum_{n=0}^{\infty} c_n(T,m_q)
\left( \frac{\mu_q}{T} \right)^n \quad ,
\label{Taylorp}
\end{equation}
where the expansion coefficients are given in terms of derivatives of
$\ln Z(T,\mu_q)$, {\it i.e.} $c_n(T,m_q) = \displaystyle{\frac{1}{n! VT^3}
\frac{\partial^n \ln Z}{\partial (\mu_q / T)^n}}$.  The series is
even in $\mu_q/T$ which reflects the invariance of $Z(T,\mu_q)$ under
exchange of particles and anti-particles. The Taylor series for the
energy density can then be obtained using the thermodynamic relation, 
$(\epsilon -3p)/T^4= T{\rm d}(p/T^4)/{\rm d}T$,
\begin{equation}
\frac{\epsilon}{T^4} = \sum_{n=0}^\infty \left(3 c_n(T,m_q) +
c'_n(T,m_q)\right) \left({\mu_q\over T}\right)^n\quad ,
\label{eps}
\end{equation}
with $c'_n(T,m_q) =T {\rm d} c_n(T,m_q)/{\rm d} T$. A similar relation holds
for the entropy density \cite{isentropic}.
The coefficients $c_n(T,m_q)$ calculated for a 
fixed value of the bare quark mass up to $n=6$ 
are shown in Fig.~\ref{fig:coefficients}.

\begin{figure}
\begin{center}
\epsfig{file=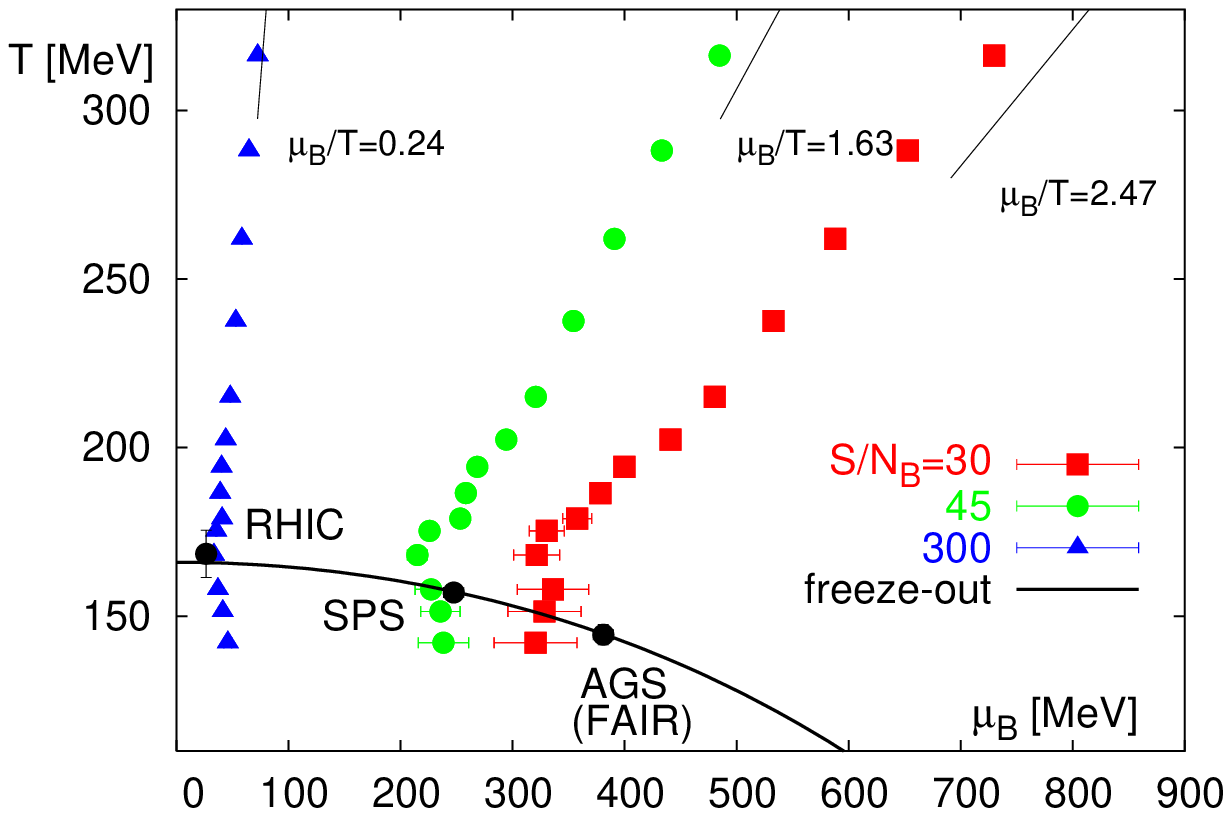,width=7.0cm}\hspace{0.2cm}
\epsfig{file=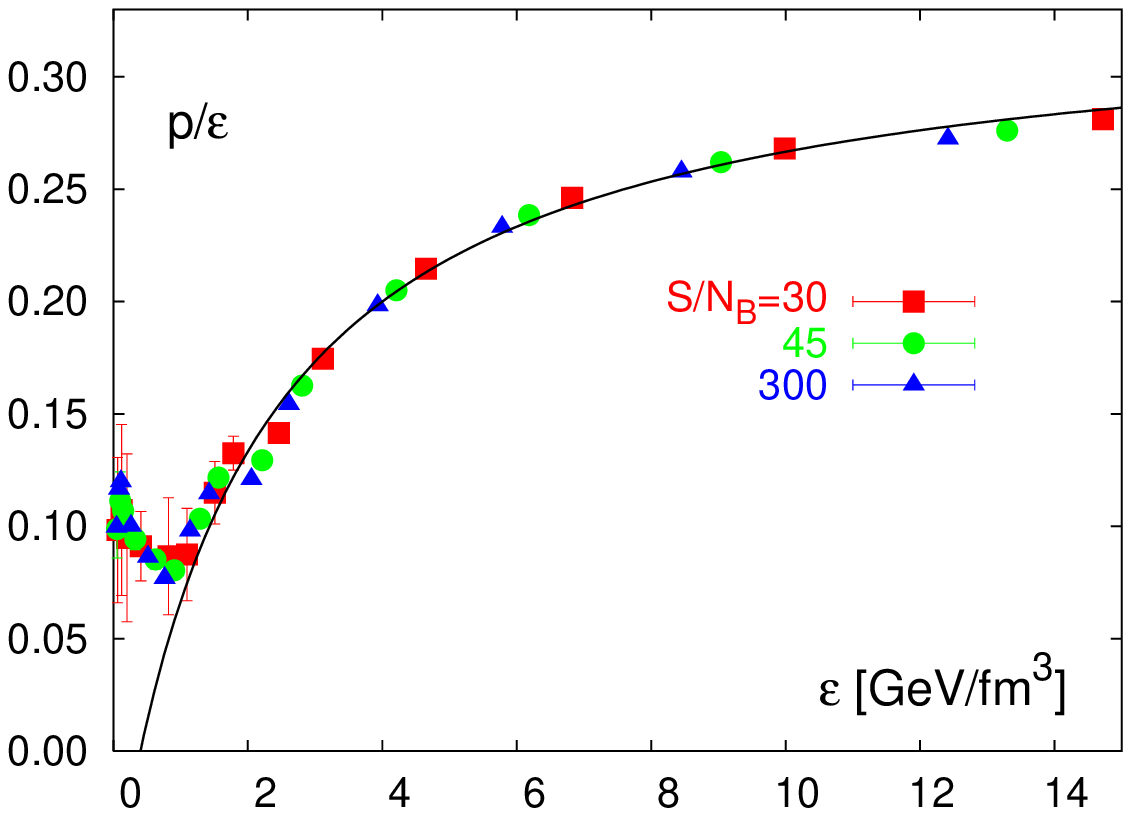, width=7.0cm}
\end{center}
\caption{Equation of state of 2-flavor QCD on lines of constant entropy per 
baryon number. The left hand figure shows three lines of constant $S/N_{\;B}$ 
in the
QCD phase diagram relevant for the freeze-out parameters determined in various
heavy ion experiments. The right hand figure shows the equation of state
on these trajectories using $T_0=175$~MeV to set the
scale. The solid curve in the right hand figure is the
parametrization of the high temperature part of the equation of state
given in Eq.~\protect{\ref{fit}}.}
\label{fig:soft}
\end{figure}

Knowing the dependence of the energy density and the pressure on the
quark chemical potential one can eliminate $\mu_q$ in favor of a 
variable that characterizes the thermodynamic boundary conditions for the 
system under consideration \cite{isentropic}. In the 
case of dense matter created in heavy ion collisions this is a combination
of entropy and baryon number.  Both quantities stay constant during the 
expansion of the system. In Fig.~\ref{fig:soft} we show the
resulting isentropic equation of state as function of temperature as 
well as energy density obtained from a $6^{th}$ order Taylor
expansion of pressure and energy density \cite{isentropic}. 
The three different entropy and baryon number
ratios, $S/N_B = 30,~45$ and $100$, correspond roughly to isentropic
expansions of matter formed at the AGS, SPS and RHIC, respectively. 
It is quite remarkable that $p(\epsilon)$ is to a good approximation
independent of $S/N_B$; for temperatures $T>T_0$, or equivalently
$\epsilon\; \gsim\; 0.8$~GeV/fm$^3$, the equation of state is well 
described by
\begin{equation}
\frac{p}{\epsilon} = \frac{1}{3}\left(1- \frac{1.2}{1+0.5\; \epsilon\; {\rm fm}^3
/{\rm GeV}}\right) \; .
\label{fit}
\end{equation}
For large energies this agrees well with a bag equation of state,
$3p= \epsilon- 4B +{\cal O}(\epsilon^{-1})$, with $B^{1/4}\simeq 260$~MeV.
Deviation from the simple bag EoS, however, become large close to $T_0$;
corrections in an expansion in terms of $\epsilon^{-1}$ exceed the
leading bag term contribution in magnitude for
for $\epsilon \lsim 5$~GeV/fm$^3$ or equivalently $T\lsim 1.5 T_0$. 

\begin{figure}
\includegraphics[width=18pc]{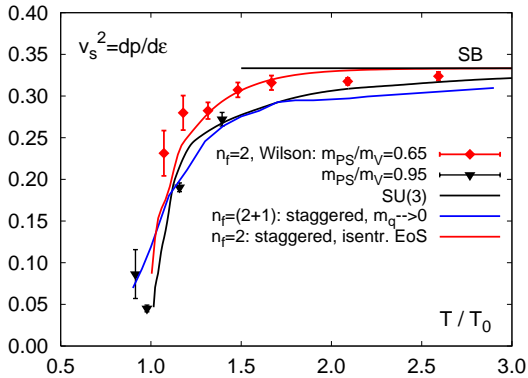}\hspace{2pc}%
\begin{minipage}[b]{18pc}\caption{\label{fig:sound}
The velocity of sound in QCD vs. temperature expressed in units
of the transition temperature $T_0$. 
Shown are results from calculations
with Wilson \cite{AliKhan} and staggered fermions \cite{aoki} as well as 
for a pure SU(3) gauge theory \cite{Boyd}.
Also shown is $v_s^2$ deduced from the isentropic equation of state,
Eq.~\ref{fit} 
\cite{isentropic}.\\[3mm]}
\end{minipage}
\end{figure}

The insensitivity of the isentropic equation of state on $S/N_B$ also
implies that the velocity of sound, $v_S = \sqrt{{\rm d} p/{\rm d}\epsilon}$,
is similar along different isentropic expansion trajectories. In fact,
the parametrization given in Eq.~\ref{fit} suggests that the velocity
of sound approaches rather rapidly the ideal gas value, $v_S^2=1/3$.
In Fig.~\ref{fig:sound} we summarize results for $v_s^2$ obtained in lattice 
calculations for a $SU(3)$ gauge theory \cite{Boyd}, 
for 2-flavor QCD with Wilson fermions \cite{AliKhan}, 
($2+1$)-flavor QCD with staggered fermions \cite{aoki} as well as from
the isentropic equation of state for 2-flavor QCD \cite{isentropic}.

\subsection{The chiral critical point}

Various model calculations \cite{Stephanov} suggest that a second order 
phase transition point (chiral critical point) exists in the QCD phase 
diagram which separates a region of first order phase transitions at high 
baryon number density and low temperatures from a cross-over region at low 
baryon number density and high temperature. Evidence for the existence
of such a critical point may come from lattice calculations at non-zero
quark chemical potential by either determining the location of Lee-Yang
zeroes \cite{Fodor1} or by determining the convergence radius of 
the Taylor series for the logarithm of the partition function which directly 
yields the pressure, $p/T= V^{-1}\ln Z$ \cite{Allton2}.

If there exists a $2^{nd}$ order phase transition point in the QCD phase 
diagram, this could be determined from an analysis of the volume 
dependence of  Lee-Yang zeroes of the QCD partition function.
In any finite
volume zeroes of $Z(V,T,\mu_q)$ only exist in the complex $\mu_q$
plane with ${\rm Im}\mu_q \ne 0$. Only for $V\rightarrow \infty$ 
some of these zeroes may converge to the real axis and will then
give rise to singularities in thermodynamic quantities. The  
relation between phase transitions and zeroes of the partition function
has been exploited using a reweighting technique to extend lattice
calculations performed at $\mu_q = 0$ to $\mu_q > 0$ \cite{Fodor1}.
Recent results based on this approach \cite{Fodor2} suggest that a critical 
point indeed exists and occurs at $\mu_B=3\mu_q \simeq 360$~MeV.  This 
estimate is about a factor two smaller than earlier estimates \cite{Fodor1}
which have been obtained on smaller lattices and with larger quark masses.
This suggests that a detailed analysis of the quark mass and volume 
dependence \cite{ejiri} still is needed to gain confidence in the analysis
of Lee-Yang zeroes.
 
The radius of convergence of the Taylor series is controlled by  
a singularity in the complex $\mu_q$ plane closest to the origin. 
It is related to the location of the critical point
only if this singularity lies on the real axis. A sufficient condition
for this is that all expansion coefficients in the Taylor series are 
positive. For temperatures below the transition temperature at $\mu_q =0$ 
this indeed seems to be the case for all expansion coefficients 
calculated so far. The first coefficient, $c_0$,
gives the pressure at $\mu_q = 0$ shown in Fig.~\ref{fig:chiral}(right)
and thus is positive for all temperatures.
This also is the case for  $c_2$, which
is proportional to the quark number susceptibility at $\mu_q = 0$
\cite{Gottlieb},
\begin{equation}
\frac{\chi_q}{T^2} = \frac{\partial^2 p/T^4}{\partial (\mu_q/T)^2} =
\sum_{n=0}^{\infty}d_n \left( \frac{\mu_q}{T} \right)^n \quad {\rm with}
\quad d_n=(n+2)(n+1)c_{n+2} \; . 
\label{chiq}
\end{equation}
As can be seen in Fig.~\ref{fig:pressure} also the next-to-leading
order coefficient, $c_4$, is strictly positive.

A new feature shows up in the expansion coefficients at ${\cal O}(\mu^6)$.
The coefficient $c_6$ is positive only below $T_0$ and changes sign in its
vicinity. If this pattern persists for higher order
expansion coefficients one may conclude that the 
irregular signs of the expansion coefficients for $T>T_0$ suggest that
the radius of convergence of the Taylor series is not related to critical
behavior at these temperatures, whereas it determines a critical point
for $T<T_0$.

Ratios of subsequent expansion coefficients provide an estimate
for the radius of convergence of the Taylor expansion,
\vspace*{-0.2cm}
\begin{equation}
\rho (T)=\lim_{n\to\infty}\rho_n\equiv
\lim_{n\to\infty}\sqrt{\left\vert {c_n\over c_{n+2}}\right\vert} =
\lim_{n\to\infty}\sqrt{\left\vert {d_n\over d_{n+2}}\right\vert} \quad .
\label{convergence}
\end{equation}
The expansion coefficients $d_n$ for the quark number susceptibility have
been analyzed recently for unimproved staggered fermions \cite{gavai} up
to $n=6$. It has been shown that an accurate determination of these expansion 
coefficients requires large physical volumes. Based on a finite volume
analysis the radius of convergence 
has been estimated from the Taylor series of the quark number
susceptibility to be $\mu_B \simeq 180$~MeV \cite{gavai}. As the expansion
coefficients $d_n$ are directly related to the expansion coefficients $c_n$
of the pressure the radius of convergence coincides in the limit $n\rightarrow
\infty$. For finite $n$, {\it i.e.} $n\simeq 6$, estimates based on the 
Taylor series for the pressure will be about 30\% higher than those based
on the Taylor series for susceptibilities. In any case, current estimates
for the location of the critical point in the QCD phase diagram differ
quite a bit and these calculations need to be refined in the future.

\section{Conclusions}

We have presented recent results obtained on the QCD equation of state at 
vanishing  and non-vanishing quark chemical potential. These calculations
now can be performed with an almost realistic quark mass spectrum. At
vanishing chemical potential one now can 
start a systematic analysis of lattice artefacts which allows 
a controlled extrapolation of lattice results on the equation
of state and the transition temperature to the continuum limit. Also 
studies of the QCD phase diagram at non-zero quark chemical potential
made significant progress in recent years. 
However, in this case much work is still needed to perform high precision 
calculations that would allow for a controlled continuum extrapolation.

\ack
This manuscript has been authored under contract number
DE-AC02-98CH1-886 with the U.S. Department of Energy.
Accordingly,
the U.S. Government retains a non-exclusive, royalty-free license to
publish or reproduce the published form of this contribution, or allow
others to do so, for U.S.~Government purposes.

\end{document}